\documentclass[
    ,final            
  ]
  {aipproc}
\layoutstyle{8x11double}

\usepackage{amssymb}

\newcommand{\bear}{\begin{eqnarray}}
\newcommand{\eear}{\end{eqnarray}}
\newcommand{\ba}{\begin{array}}
\newcommand{\ea}{\end{array}}
\newcommand{\nn}{\nonumber}

\begin{document}

\title{BPS D-branes from an Unstable D-brane
\footnote{Proceeding of PASCOS 2005, Gyeongju, Korea, May 30-June
4, 2005. Talk was given by O-K. Kwon.}}


\author{Chanju Kim}{
  address={Department of Physics, Ewha Womans University,
Seoul 120-750, Korea}
}

\author{Yoonbai Kim}{
  address={BK21 Physics Research Division and Institute of
Basic Science,\\ Sungkyunkwan University, Suwon 440-746, Korea}
}

\author{Hwang-hyun Kwon}{
  address={BK21 Physics Research Division and Institute of
Basic Science,\\ Sungkyunkwan University, Suwon 440-746, Korea}
}

\author{O-Kab Kwon}{
  address={BK21 Physics Research Division and Institute of
Basic Science,\\ Sungkyunkwan University, Suwon 440-746, Korea}
}

\author{\\Chong Oh Lee}{
  address={Department of Physics, Chonbuk National University,
Chonju 561-756, Korea}
}

\begin{abstract}

We search for exact tachyon kink solutions of DBI type
effective action describing an unstable D-brane with worldvolume
gauge field turned in both the flat and a curved background. There
are various kinds of solutions in the presence of electromagnetic
fields in the flat space, such as periodic arrays, topological
tachyon kinks, half kinks, and bounces. We identify a BPS object,
D($p$-1)F1 bound state, which describes a thick brane with string
flux density. The curved background of
interest is the ten-dimensional lift of the Salam-Sezgin vacuum
and, in the asymptotic limit, it approaches ${\rm R}^{1,4}\times
{\rm T}^2\times {\rm S}^3$. The solutions in the curved background
are identified as composites of lower-dimensional D-branes
and fundamental strings, and, in the BPS limit, they become
a D4D2F1 composite wrapped on ${\rm R}^{1,2}\times {\rm T}^2$
where ${\rm T}^2$ is inside ${\rm S}^3$.

\end{abstract}

\maketitle


\section{Introduction and Summary}

Study of the dynamics of unstable D-brane has been realized by condensation
of tachyonic mode living on the brane~\cite{Sen:2004nf}.
After A. Sen wrote down the boundary state describing decay of
unstable D-branes in boundary conformal field
theory~(BCFT)~\cite{Sen:2002nu} and suggested the corresponding tachyon
effective action~\cite{Sen:2002an}, such observations enables us to study
the decay of the unstable D-branes in the spatially homogeneous background,
say, the rolling tachyon. On the other hand,
spatial inhomogeneity has been another important issue, particularly
in the form of tachyon solitons which form BPS
and non-BPS D-branes of various codimensions.

It is well known that the tachyon effective field theory~(EFT) correctly
captures some aspects of the tachyon condensation in the low energy
limit of open string theory. Particularly, tachyon kink solutions are
effectively described in EFT. The purpose of this note is to review
the tachyon kink solutions with electromagnetic fields
under a runaway tachyon potential, $1/\cosh(T/\sqrt{2})$, in flat
and a curved background, and summarizes the resulting BPS objects
on the unstable D-branes.

In flat space, the obtained static kink configurations for pure tachyon
fields are either singular solutions~\cite{Sen:2003tm}
or an array of regular
kink-anti-kink~\cite{Lambert:2003zr,Brax:2003rs,Kim:2003in}.
Once constant DBI-type electromagnetic fields are turned on,
there are additional five non-trivial regular solutions.
Specifically the solutions include
two types of topological kinks, bounce, half kink,
and hybrid of two half kinks~\cite{Kim:2003in}.
When the pure electric field along the inhomogeneous direction
 is less than or equal to 1, corresponding BCFT solutions are also obtained
in Ref.~\cite{Sen:2003bc}.
Remarkably, in the critical limit of the
electric field, $|{\bf E}|=1$, the resulting solution represents a
D($p$-1)F1 bound state and is identified as the thick BPS-brane
with string flux density. The thickness can be adjusted
by the strength of the string flux density.

We attempt to extend the analysis of unstable D-brane to the case of a
curved bulk background and find tachyonic kink
solutions~\cite{Kim:2004xk,Kim:2005he}. Similar problems were
considered in Ref.~\cite{Kluson:2004yk}.
The background of our consideration
is ${\rm R}^{1,4}\times {\rm T}^2 \times {\rm S}^3$
with non-vanishing NSNS $B$-field, which is the asymptotic limit of the ten-dimensional
embedding~\cite{Cvetic:2003xr} of the supersymmetric vacuum,
${\rm R}^{1,3}\times {\rm S}^2$, of the Salam-Sezgin model~\cite{Salam:1984cj}.
We obtain exact tachyon kink solutions on a non-BPS D5-brane
whose worldvolume lies on ${\rm R}^{1,2}\times {\rm S}^3$.
The obtained solutions describe the codimension-one branes on the non-BPS
D-brane. In the thin limit of the solution,
it becomes a BPS object with string flux density when a constant
magnetic field $h$ is goes to zero and forms a D4D2F1 bound state.

\section{Generalities}

In this paper we consider the DBI-type effective action for
tachyon field which couples to abelian gauge field
on an unstable Dp-brane in general backgrounds,
\bear \label{dbia}
S = \int d^{p+1} x {\cal L}=
- {\cal T}_p \int d^{p+1} x\,e^{-\Phi} V(T) \sqrt{-X},
\eear
with
\bear
X &\equiv& \det X_{\mu\nu}
=\det (g_{\mu\nu}
      + \partial_\mu T \partial_\nu T + {\cal F}_{\mu\nu}),
\nn \\
{\cal F}_{\mu\nu} &=& B_{\mu\nu} + F_{\mu\nu},
\nn \\
F_{\mu\nu}&=& \partial_\mu A_\nu - \partial_\nu A_\mu, \quad (\mu,
\nu = 0,1, \cdots, p), \nn \eear where ${\cal T}_p$ is the tension
of the unstable Dp-brane, $\Phi$ is the dilaton, $B_{\mu\nu}$ is
the induced anti-symmetric tensor. We neglected the transverse
scalars which are irrelevant in our discussion. Equations of
motion for the tachyon $T$ and gauge field $A_\mu$ are given by
\bear & & \nabla_\mu\left[\frac{\gamma_p}{\sqrt{-g}}
\;C^{\mu\nu}_{\rm S}\;
\partial_\nu T\right] +\frac{\sqrt{-X}}{\sqrt{-g}}\;{\cal T}_p
e^{-\Phi}\frac{d V}{d T} = 0,
\label{te} \\
& &\nabla_\mu\left[ \frac{\gamma_p}{\sqrt{-g}} \;C^{\mu\nu}_{\rm A}
\right] = 0, \label{ge}
\eear
where $C^{\mu\nu}_{\rm S}$ and $C^{\mu\nu}_{\rm A}$ are the symmetric and
anti-symmetric parts of the cofactor, $C^{\mu\nu}$, of the matrix
$(X)_{\mu\nu}$, and we define
\bear\label{gamp}
\gamma_p \equiv \frac{{\cal T}_pe^{-\Phi}V}{\sqrt{-X }}.
\eear
Conservation of energy-momentum is given by
\bear\label{cem}
\nabla_\mu T^{\mu\nu} = \nabla_\mu\left[ \frac{\gamma_p}{\sqrt{-g}}\;
C^{\mu\nu}_{\rm S}  \right] =  0,
\eear
where $T^{\mu\nu}$ is the energy-momentum tensor.
For the tachyon potential $V(T)$ that is symmetric under
$T\,\to\, -T$ in type IIA and IIB superstring theories,
any runaway potential with $V(0) =1$ and $V(\pm\infty)=0$ is allowed
for the existence of D-brane configuration of our interest, that
is consistent with universal behavior in tachyon
condensation~\cite{Sen:2003tm}. Here we assume
a specific form in order to obtain exact solutions
\bear\label{tpot}
V(T) = \frac{1}{\cosh \left(\frac{T}{\sqrt{2}}\right)},
\eear
which was derived in the open string theory using
perturbation around half S-brane~\cite{Kutasov:2003er}.

\section{Tachyon Kinks in flat space}

In this section we analyze the DBI-type effective action (\ref{dbia})
with vanishing dilaton $\Phi$ and anti-symmetric field $B_{\mu\nu}$
in flat ($p$+1)-dimensions ($g_{\mu\nu} = \eta_{\mu\nu}$), and classify all
possible static tachyon kinks~(See Ref.~\cite{Kim:2003in}).
To describe the tachyon kink solutions we assume that the tachyon
and gauge field strengths depend only on one spatial direction, $x$,
\bear \label{ansz}
T = T(x), \qquad F_{\mu\nu} = F_{\mu\nu}(x).
\eear
Then the effective action (\ref{dbia}) is simplified as
\bear
S= - {\cal T}_p \int d^{p+1} x\, V(T)\sqrt{\beta_p - \alpha_{px} T'^2},
\eear
where $T' = \partial_x T$,
$\alpha_{px}$ is 11-component of the cofactor $C^{\mu\nu}$,
and $\beta_p = -\det (\eta_{\mu\nu} + F_{\mu\nu})$.

Bianchi identity for the abelian gauge field strength,
$dF = 0$, the gauge field equation (\ref{ge}), and conservation of
energy-momentum (\ref{cem}) determine the system
completely.
As expected the resulting solutions are consistent with the tachyon
equation (\ref{te}). What we obtain is as follows. All components of
the gauge field strength are constants, and $T^{11}$ which is written as
$T^{11} = \gamma_p\alpha_{px}$ is a constant of motion.
Now the remaining equation is constancy of $T^{11}$ where the expression
of $T^{11}$ is reexpressed by a
first-order differential equation for the tachyon field,
\bear\label{mste}
{\cal E}_p = \frac{1}{2} T'^2 + U_p(T),
\eear
where ${\cal E}_p = \beta_p/2\alpha_{px}$ and
$U_p = \alpha_{px} \left[{\cal T}_p V(T)\right]^2/2 (T^{11})^2$.
We can classify all kink solutions in terms of three parameters
($\alpha_{px}$, $\beta_p$, $T^{11}$). For the tachyon potential
(\ref{tpot}) all the codimension-one tachyon solitons are given by
exact solutions~\cite{Kim:2003in}.

When  $\alpha_{px}$ is negative, $U_p$ is turned upside down and has the
minimum value, $\alpha_{px}{\cal T}_p^2/2(T^{11})^2$, at $T=0$ and the
maximum value, $0$,
at $T=\pm\infty$ due to the runaway property of the tachyon potential
(\ref{tpot}).
Three types of the exact solutions of Eq.~(\ref{mste}) are
\bear
\sinh\left(\frac{T(x)}{\sqrt{2}}\right)=~~~~~~~~~~~~~~~~~~~~~~~
~~~~~~~~~~~~~~~~~~~~~~~~~~~~~~~
\eear
\bear
\left\{
\begin{array}{cclc}
\sqrt{u^{2}-1} \sin\left(x/\zeta\right)
&\mbox{for}&0<\beta_{p}<\tilde u^2&(\mbox{i})\\
ux/\zeta
&\mbox{for}&\beta_{p}=0 &(\mbox{ii})\\
\sqrt{u^{2}+1}\sinh \left(x/\zeta\right)
&\mbox{for}&\beta_{p}<0 &(\mbox{iii})
\end{array}\right.,\nn
\eear where \bear\nn u^{2}=\left|\frac{{\cal
T}_{p}^{2}\alpha_{px}^{2}} {\beta_{p}(T^{11})^{2}}\right|, \quad
\tilde u^2 = \beta_p u^2, \quad \zeta=\sqrt{\left|
\frac{2\alpha_{px}}{\beta_{p}}\right|}\, . \eear (i) is an array
of kink-anti-kink which is interpreted as an array of
D($p$-1)$\bar {\rm D}$($p$-1) ( and  D($p$-1)${\rm F}1- \bar {\rm
D}$($p$-1)F1), and is an unique nontrivial solution in the pure
tachyon case~\cite{Lambert:2003zr}. (ii) and (iii) are the
topological kinks connecting two vacua $T=\pm\infty$, and
interpreted as a single D($p$-1)F1 bound state. For (i) and (ii),
BCFT calculation confirms this interpretation~\cite{Sen:2003bc}.

When $\alpha_{px}$ is positive, there are three more types of non-trivial
solutions
\bear\label{palp}
&&\sinh\left(\frac{T(x)}{\sqrt{2}}\right)=
 \\
&&\left\{
\begin{array}{cclc}
\sqrt{u^{2}-1} \cosh\left(x/\zeta\right)
&\mbox{for}& 0<\beta_{p}<\tilde u^2&  (\mbox{iv})\\
\exp \left(x/\zeta\right) &\mbox{for}&
\beta_{p}=\tilde u^2 & (\mbox{v})\\
\sqrt{1-u^2} \sinh\left(x/\zeta\right)
&\mbox{for}&\beta_{p}> \tilde u^2 & (\mbox{vi})\nn
\end{array}
\right..
\eear
These solutions are interpreted as bounce for (iv), half-kink
for (v), and hybrid of two half-kinks for (vi).
The functional form of these three solutions in Eq.~(\ref{palp})
coincides with the exact rolling tachyon solution in DBI-type
tachyon effective action~\cite{Kim:2003he}.
However, these solutions are not yet obtained in other descriptions
of the system theories, such as BCFT and boundary string field
theory~(BSFT).

In relation to BPS nature, the single topological kink (ii)
saturates BPS-type bound with thickness for $T^{11}=\Pi^1\ne 0$
case, i.e., the energy of this object consists of the string
charge and RR-charge of the lower-dimensional D($p$-1)-brane
exactly.

\section{Tachyon Kinks in a Curved Background}

Since the DBI action can describe the low energy dynamics of
D$p$-brane in a curved background as well, we consider the
DBI-type tachyon effective action on the curved
background~\cite{Kim:2004xk, Kim:2005he,Kluson:2004yk}. This
section is based on the Ref.~\cite{Kim:2005he}, and we
will study tachyon kink solution in the large dilaton limit of the
ten-dimensional lift of Salam-Sezgin vacuum on ${\rm
R}^{1,3}\times {\rm T}^2\times {\rm R}_\rho\times {\rm S}^3$
described by~\cite{Cvetic:2003xr} \bear ds^{2}&=& dx_6^{2}
+4R^{2}d\rho^{2} + du^2
\nn \\
&&+ \sin^2 \left(\frac{u}{R}\right)\, dv^2
 +\left[dw + \cos
\left(\frac{u}{R}\right)\, dv\right]^{2},
\nn\\
B &=& -\cos \left(\frac uR \right) dv \wedge dw ,
\nn \\
\Phi &=& -\rho ,\label{liftss}
\eear
where $R$ is the radius of ${\rm S}^3$ which is parametrized by three
coordinates ($u$, $v$, $w$), $B$ is the non-vanishing
NS-NS two-form field on ${\rm S}^3$, and $\Phi$ is the dilaton field.
It is interesting to notice that the local geometry of the background
(\ref{liftss}) is nothing but the NS5-brane near horizon geometry.
However, there is a difference in that string coupling constant goes to
zero in the asymptotic limit of the background, while it blows up
in the throat region of the NS5-brane. Thus in this background (\ref{liftss}),
it is valid to study non-BPS D-branes in terms of DBI-type effective
theory (\ref{dbia}).
Note that ${\cal F}_{\mu\nu}$ defined by
${\cal F}_{\mu\nu} =F_{\mu\nu} + B_{\mu\nu}$ is gauge-invariant
on the worldvolume of D-brane.

We consider an unstable D5-brane on ${\rm R}^2\times {\rm S}^3$ with
the coordinates ($x,z,u,v,w$) where $(x,z)$ is two of the spatial
coordinates of ${\rm R}^{1,3}$. Similar to the case of the flat
space in the previous section, we assume the same tachyon potential
(\ref{tpot}). The compactification scale $R$ in the Eq.~(\ref{liftss})
is assumed to be identical to the self-dual radius $\sqrt{2}$
in the tachyon potential (\ref{tpot}).

From now on, we will study static solutions of the tachyon effective
action (\ref{dbia}) under the ansatz
\bear\label{anszc}
&&T=T(u,w),
 \\
&&F_{0z}=E_{z}(u,w),~F_{vz}=\alpha(u,w),
~F_{vw}=h(u,w),\nn
\eear
where we assume other components of the gauge field
strength $F_{\mu\nu}$ vanish.
Using the Bianchi identity for the two-form field
on the unstable D-brane,
$d{\cal F} = H$, where $H$ is field strength of the anti-symmetric
tensor field, we obtain
\bear\label{bian}
 E_{z}={\rm constant},\quad \alpha={\rm constant},\quad h=h(w).
\eear
Inserting the Eqs.~(\ref{liftss}), (\ref{anszc}), and (\ref{bian})
into the equations of motion (\ref{te}) -- (\ref{ge}),
we find that the tachyon field $T$ is a function of either $u$ or
$w$ but not both, and $h$ should be a constant, i.e.,
\bear
&&\partial_u T\partial_w T =0,
\nn \\
&&h = {\rm constant},
\eear
and the dilaton field $\Phi$ is decoupled.

We first consider the case that $T$ depends only on $u$.
Then we can find an exact solution
\bear \label{tub}
\sinh\left(\frac{T(u)}{\sqrt{2}}\right)
=\pm \sqrt{\left(
\frac{{\cal T}_5}{\alpha\gamma_5}\right)^{2}-1}
\,\cos \left(\frac{u}{\sqrt{2}}\right),
\eear
where $\gamma_5$ was defined in Eq.~(\ref{gamp}).
From the analysis of the physical quantities for the solution
(\ref{tub}), e.g., energy-momentum tensor $T^{\mu\nu}$ and electric
flux density $\Pi^i$,
we notice that the solution represents a dimensionally reduced
configuration. From the background metric in Eq.~(\ref{liftss}),
 the configuration  spans ${\rm T}^{2}$ in the
three-sphere:
\bear\label{dst2}
ds_{{\rm S}^{3}}&=&du^{2}
+ \sin^2 \left(\frac{u}{\sqrt{2}}\right)\, dv^2 +\left[dw + \cos
\left(\frac{u}{\sqrt{2}}\right)\, dv\right]^{2}
\nn \\
&&~\stackrel{\frac{u}{\sqrt{2}}=\frac{\pi}{2}}{\longrightarrow}~
ds_{{\rm T}^{2}}=dv^2 + dw^2 .
\eear

To identify the RR-charge of the resulting lower dimensional D-brane
we take into account the Wess-Zumino term for the unstable
D-brane~\cite{Sen:2003tm,Okuyama:2003wm},
\begin{equation} \label{wzterm}
S_{\rm WZ} = {\cal T}_{5}\int V(T)\, dT \wedge C_{{\rm RR}}
\wedge e^{F+B}.
\end{equation}
For the solution (\ref{tub}) in the thin limit $\gamma_5\to 0$,
the Wess-Zumino term (\ref{wzterm}) is reduced to
\bear
S_{\rm WZ} &=& \mp \pi \sqrt{2} {\cal T}_5 \int \left[ C_{(5)}
+ \alpha\,C_{(3)} \wedge dz\wedge dv \right.
\nn \\
&&\left. \hskip 1.5cm + h\,\tilde C_{(3)}\wedge dv\wedge dw \right],
\eear
where we omitted the terms containing a RR-form wedged to $dt$,
which are irrelevant in the interpretation of the lower-dimensional
D-brane. Thus the resulting configuration consists of a D4-brane
stretched along ${\rm R}^2 \times {\rm T}^2$ with coordinates
$(x,z,u,w)$ and RR-charge ${\cal T}_4 = \sqrt{2}\pi {\cal T}_5$ and
two D2-branes with charges per unit area, $\sqrt{2} \pi {\cal T}_5\alpha$
and $\sqrt{2} \pi {\cal T}_5 h$,  spanned by the worldvolume coordinates
$(t,x,w)$ and $(t,x,z)$, respectively. In addition, there are
fundamental strings with flux on cylinder ${\rm R}\times {\rm S}_1$
of $(z,w)$.

In order to study the BPS nature of the solution, we now investigate the
energy-momentum tensor.
For the solution (\ref{tub}) to describe a BPS object,
it is required that the pressure in $u$-direction, $T^{uu}$,
and the off-diagonal stress component between two D2-branes,
$T^{zw}$, should vanish, which are given by
\bear
T^{uu} &=& -\frac{\gamma_5 \alpha^2}{\sin (u/\sqrt{2})},
\nn \\
T^{zw} &=& \frac{\alpha}{\sin (u/\sqrt{2})}
\left[ h-\cos\left(\frac{u}{\sqrt{2}}\right) \right] \Sigma(u)
\eear
with
\bear \label{sig}
\Sigma(u) =\frac{ {\cal T}_5^2/\gamma_5 \alpha^2}{[
(\frac{{\cal T}_5}{\gamma_5  \alpha})^2 - 1 ] \cos^2(\frac u{\sqrt{2}}) + 1}.
\eear
To obtain this vanishing condition we should take the thin limit,
$\gamma_5 \to 0$, with $\pi/\sqrt{2}$ and $h=0$.
Then we have single D2-brane along the direction ($x$, $w$) and
an electric flux along the $z$-direction. This solution is expected
to saturate a BPS bound. The energy per unit area of coordinates
($x$, $w$) takes the form
\bear \label{bpssr}
\frac{{\cal H}}{\int dx\, dw}
&=& \int dz\,du\,dv\,\Pi_z
+ \sqrt{2}\pi {\cal T}_5 \alpha \int dz\,dv\, \nonumber \\
&=& Q_{\rm F1} + Q_{\rm D2},
\eear
where $Q_{\rm F1}$ is the total charge of fundamental strings along
$z$-direction and $Q_{\rm D2}$ the total charge of D2-brane stretched along
($x$, $w$)-direction on the area $\int\,dz\,dw$.

In summary, when $h=0$, the solution produces a BPS
D4D2F1-composite in the thin
limit, $\gamma_5 \rightarrow 0$. It consists of
the D4-brane wrapped on R$^{2}\times$T$^2$,  the tubular
D2-brane with the coordinates $(x,w)$, and the fundamental
strings stretched along $z$-direction.

For the case $T=T(w)$ we also obtain an exact solution
similar to the previous solution (\ref{tub}) and investigate again the BPS
nature in the thin limit $\gamma_5 \to 0$ which satisfies the condition
$h=0$ automatically in this case. This BPS object is proven to describe
the same configuration as the BPS solution in the case of $T=T(u)$.

\begin{theacknowledgments}

This work was supported by Korea Science and Engineering Foundation
(KOSEF R01-2003-000-10229-0 for C.K. and R01-2004-000-10526-0 for C.L.),
and is the result of research activities (Astrophysical Research
Center for the Structure and Evolution of the Cosmos (ARCSEC))
supported by Korea Science $\&$ Engineering
Foundation~(Y.K., H.K., and O.K.).

\end{theacknowledgments}


\end{document}